\documentclass[review]{elsarticle}

\usepackage{amsmath}
\usepackage{amsfonts}

\usepackage{tikz}
\usetikzlibrary{arrows}

\usepackage{floatrow}

\usepackage{url}

\usepackage{caption,subcaption}

\usepackage{lineno}

\begin{document}

\title{Inference of population-level disease transmissibility from
  household-structured symptom onset data}


\author[1]{P. G. Ballard\corref{cor1}}
\ead{peter.ballard@adelaide.edu.au}

\author[2]{A. J. Black}
\ead{andrew.black@adelaide.edu.au}

\author[3]{J. V. Ross}
\ead{joshua.ross@adelaide.edu.au}

\cortext[cor1]{Corresponding author}

\fntext[fn1]{School of Mathematical Sciences,
  The University of Adelaide, Adelaide SA 5005, AUSTRALIA.}

\begin{abstract}
  First Few X (FFX) studies
  collect household-stratified data in the early stages of a pandemic,
  in order to infer severity and transmissibility of an emerging disease.
We present a Bayesian method to approximately infer population-level transmissibility for the first time from such data;
previous studies have only inferred household-level transmissibility.
To do this we perform the inference at two levels,
assuming one transmission rate parameter for within-household infection,
and another transmission rate parameter for infection between different households.
We use a simplifying assumption:
that between-household infections always occur in naive households;
while still performing full joint inference on the within-household infection parameters.
In addition, a novel technique is used to remove systematic bias when the number of new infections per day is growing or decaying,
as is common in real outbreaks,
without the need for 
contemporaneous estimates of the serial interval.
The method is validated on simulated data and is shown to perform well,
even when the number of infected households is relatively
small.
\end{abstract}


\begin{frontmatter}
\end{frontmatter}



\section{Introduction}\label{intro}

The threat from pandemics is ongoing \cite{WHOFLU:2017},
and this threat has been underscored by the outbreak of COVID-19 \cite{WHO2021}.
An important part of dealing with this threat is the ability to infer the characteristics of an emerging strain of infection,
in particular its transmissibility and severity, which are strong determinants of impacts of the disease \cite{McCaw2013,Reed2013,Riley2015}.

The accurate inference of transmissibility is particularly challenging,
due to the inherently heterogeneous mixing present in larger populations.
Models with two levels of mixing, or so-called household models \cite{Ball1997, Ball2002b},
are an established way to capture some of this heterogeneity \cite{House2008, Ross2010, Hilton2019}.
These assume stronger mixing within a household, and weaker mixing in the overall population between individuals in different households \cite{Ball1997,Ross2010,Black2013}.
Households are a convenient unit to monitor because contacts are easy to track;
and although different rates of mixing have been shown to exist within households \cite{Mossong:2008,Geard:2013},
within-household mixing can be assumed to be homogeneous as a first order approximation.

As one part of the response to the threat of a pandemic,
governments around the world have invested in advanced plans for data collection in its early stages, so-called
``First Few X'', or FFX, studies \cite{AHMPPI2014,HPA:2009,McLean:2010,Gag-Laf:2012}. 
As this data is already at the household level, it is natural to consider its use in
the inference of transmissibility in a two-level model of the entire population.
Previous work has shown how mechanistic models can be used to accurately infer transmissibility \emph{within a household} using FFX data \cite{Black2017}.
In this paper, we build on this and demonstrate how FFX data can also be used to infer
\emph{population level} transmissibility.
The method presented here is effective even if some cases are asymptomatic and hence some infected households are not observed.

The method is a three-step process.
The first step, the inference of within-household parameters,
uses similar methods to our previous work \cite{Black2017,Black2018},
and so is not described in detail in this paper.

The second step infers the between-household component of transmissibility.
This uses the within-household parameters inferred during the first step.
By using these inferred within-household parameters,
we avoid the need to contemporaneously estimate the generation time or serial interval
(for instance via contact tracing),
as is used in established methods \cite{Cori2013,Thompson2019}
but has its difficulties.\footnote{For instance:
  difficulties in determining the chain of transmission \cite{Griffin2020};
  difficulties in obtaining enough data points to be a representative sample \cite{Griffin2020};
  subjects having faulty recollection of events \cite{Griffin2020};
  or failing to account for the fact that the average serial interval will be shorter
  within a household, due to exhaustion of susceptibles \cite{Liu2018}.
  }

The third and final step of our method is to combine
the within-household and between-household parameters to
obtain the posterior distribution of whichever reproduction numbers are desired.
Those most commonly reported are 
$R_0$ (the mean number of secondary infections per infectious individual, in an otherwise uninfected population),
or $R_\text{eff}$ (the mean number of secondary infections per infectious individual, during an outbreak).
There is some question over what is the most meaningful way to precisely define
$R_0$ or $R_\text{eff}$ in a two-level model \cite{Ball1997,Pellis2012,Goldstein2009},
or even whether the between-household reproduction number $R_*$
(the mean number of households secondary households infected per infectious household)
\cite{Ross2010} is a more useful measure of transmissibility.
In any case, any of these reproduction numbers can be derived from the
within-household and between-household parameters.
It is also possible to infer the posterior distribution of the growth rate,
which can then be used for forecasting, if exponential growth or decay is
assumed \cite{Koczkodaj2020}.

Due to 
the small amount of FFX data in Australia during the COVID-19 pandemic
-- partly due to 
the fortunate situation
of there not being a large number of COVID-19 cases --
this study uses synthetic data.
This use of synthetic data allows us to make a number of simplifying assumptions.
We present these assumptions, and the details of the data and model, in Section \ref{model}.
We outline the method and its theory in Section \ref{method},
including two alternative methods in Sections \ref{indiv} and \ref{growth}.
We present sample results in Section \ref{impl} and a conclusion in Section \ref{conclusion}.

\section{Data and Model}\label{model}

\subsection{Data}

For every household which has had a symptomatic infection,
the number of individuals in that household is recorded,
as well as a daily record of the number of newly symptomatic individuals each day.
This corresponds to the Australian pandemic response plan \cite{AHMPPI2014},
which specifies that this daily data should be obtained from all of the FFX infected households.
We assume that every symptomatic case will be detected and recorded
and that the day of first symptoms can be accurately determined.
We also assume that symptoms always correspond to the disease of interest.


\subsection{Model}


The model consists of a set of households,
each represented by a stochastic compartmental model (Section \ref{within_model}),
with the addition of between-household infection events (Section \ref{between_model}).

As synthetic data is used, we have made some simplifying assumptions.
We assume that there is effectively an infinite supply of naive households,
meaning that the spread between households can be approximated as a branching process \cite{Walker2017},
which is a reasonable assumption in a population containing a large number of
households. Following on from this,
we also assume that every between-household infection infects a naive household;
in other words, that no household contains two or more individuals who have been infected
by someone outside of the household. 

We also assume that, apart from the initial ``seed'' infection into a population,
there are no further infections from outside the population.
This is a less realistic assumption, and the tracing associated with the spread of COVID-19
has shown that populations continually receive external infections, at some rate.
Some analyses have this external infection as a separate rate,
splitting infection into ``local'' and ``imported'' \cite{Thompson2019}.
We ignore this factor, though in principle it would not be difficult to account for it.

\subsubsection{Within-household model}\label{within_model}

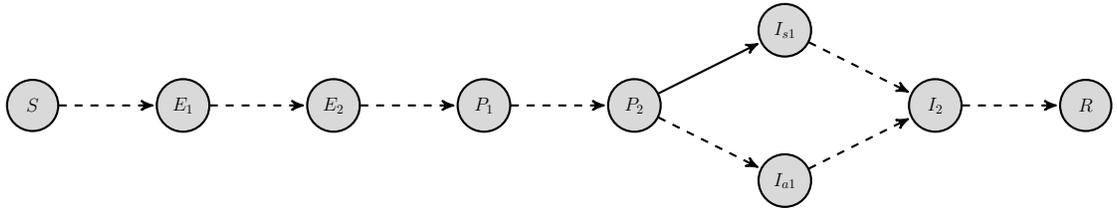
\begin{figure}[ht]
  \caption{Individuals' compartments in the COVID-19 within-household model.
    Possible changes in an individual's compartment are shown as solid or dashed lines,
    with the only observable change ($P_2\rightarrow I_{s1}$) a solid line.
    Between-household infections are not shown in this diagram.
  }
\label{fig:modelfig} 
\begin{tikzpicture}[->,>=stealth',shorten >=1pt,auto,node distance=2.5cm,
    thick,main node/.style={circle,text width=1cm,text centered,
    fill=gray!30,draw,font=\sffamily\Large\bfseries},
    scale=0.5, every node/.style={transform shape}]


  \node[main node] (S) at (-8, 0) {$S$};
  \node[main node] (E1) at (-4, 0) {$E_1$};
  \node[main node] (E2) at (0, 0) {$E_2$};
  \node[main node] (P1) at (4, 0) {$P_1$};
  \node[main node] (P2) at (8, 0) {$P_2$};
  \node[main node] (Is1) at (12, 2) {$I_{s1}$};
  \node[main node] (I2) at (16, 0) {$I_{2}$};
  \node[main node] (Ia1) at (12, -2) {$I_{a1}$};
  \node[main node] (R) at (20, 0) {$R$};

  \path[every node/.style={font=\sffamily\small}]
  (S) edge [style=dashed]  node {} (E1) 
  (E1) edge [style=dashed]  node {} (E2)
  (E2) edge [style=dashed]  node {} (P1)
  (P1) edge [style=dashed]  node {} (P2)
  (P2) edge [color=black] node {} (Is1) 
  (P2) edge [style=dashed]  node {} (Ia1)
  (Is1) edge [style=dashed]  node {} (I2)
  (Ia1) edge [style=dashed]  node {} (I2)
  (I2) edge [style=dashed]  node {} (R);
\end{tikzpicture}
\end{figure}

The methods in this paper can be adapted to a number of different compartmental household models,
but we focus here on a single model to illustrate the methods:
a model designed to emulate the progress of a COVID-19 infection.
The compartments, which are the possible states of an individual, are illustrated in Figure \ref{fig:modelfig}.
The process is modelled as a continuous-time Markov chain (CTMC)
and the events, rates and parameters are given in Table \ref{tab:modeltab}.
The model is an extension of the common SEEIIR model
with partial observation
\cite{Black2017, 
  Mitchell2016, 
  Price2020}; 
in that it has two ``pre-symptomatic'' compartments
to account for the time in which
an infected individual is infectious before symptoms appear, as occurs with COVID-19.
The model assumes that infectiousness is equal in all of the ``infectious'' compartments
($P_1$, $P_2$, $I_{s1}$, $I_{a1}$ and $I_2$).

Individuals are symptomatic with probability $p_s$,
and the $I_{a1}$ compartment denotes cases which are never symptomatic.

\begingroup
\renewcommand{\arraystretch}{1.5}
\begin{table}[H]
\begin{tabular}{llll}
\hline
Description & Transition & Rate & Observed \\
\hline
Exposed to infection & $S\rightarrow E_1$ & $\beta SI/(m-1)$ & no \\
Exposed state progresses & $E_1 \rightarrow E_2$ & $2E_1/t_E $ & no  \\
Becomes infectious, pre-symptomatic & $E_2\rightarrow P_1$ & $2E_2/t_E $ & no \\
Pre-symptomatic state progresses & $P_1 \rightarrow P_2$ & $2P_1/t_P$ & no  \\
Becomes infectious with symptoms & $P_2 \rightarrow I_{s1}$ & $2P_2\,p_s/t_P$ & yes  \\
Infectious state progresses & $I_{s1} \rightarrow I_{2}$ & $2I_{s1}/t_I$ & no  \\
Becomes infectious, no symptoms & $P_2 \rightarrow I_{a1}$ & $2P_2 (1-p_s)/t_P$ & no  \\
Infectious state progresses & $I_{a1} \rightarrow I_{2}$ & $2I_{a1}/t_I$ & no  \\
Recovery & $I_{2} \rightarrow R$ & $2I_{2}/t_I$ & no  \\
\hline
\end{tabular}
\caption{Within-household transition rates for the COVID-19 
  model. 
  In the ``Rate'' column, by a slight abuse of notation,
  the compartment name refers to the number of individuals in that compartment.
  $\beta$ is the within-household infection rate parameter;
  $I = (P_1+P_2+I_{s1}+I_{a1}+I_{2})$
  is the total number of infectious individuals in the household;
  $m$ is the number of individuals in a household;
  $t_E$, $t_P$ and $t_I$ are the mean times in
  the exposed (but not infectious),
  pre-symptomatic (and infectious)
  and symptomatic states respectively;
  and $p_s$ is the probability of an infectious individual becoming symptomatic.
  }
\label{tab:modeltab}
\end{table}
\endgroup

\subsubsection{Between-household model}\label{between_model}

For the process of between-household infection, we assume that the population of naive
(that is, fully susceptible) households is large,
so new infections can be modelled as a branching process.
Thus the rate at which new households become infected is given by $\alpha$,
the rate of between-household infections per infectious individual,
multiplied by the total number of infectious individuals
(that is, the sum of individuals in the compartments $P_1$, $P_2$, $I_{s1}$, $I_{a1}$ and $I_{2}$)
across all households.
As the supply of susceptible households is effectively infinite, 
this term does not need to be reduced as the outbreak progresses,
unlike the within-household infection rate.

Since it is assumed that no household is infected from another household more than once,
any newly infected household initially has
one individual in the $E_1$ compartment, and all other individuals in the $S$ compartment.
However the household is not observed, if it is observed at all,
until one of the individuals in the household becomes symptomatic;
that is, until a $P_2 \rightarrow I_{s1}$ event occurs for one of the individuals in the household.

\section{Method}\label{method}


We infer the reproduction numbers by a three-step process:
(1) treat households as independent populations, and infer the within-household model parameters; 
(2) treat households as units, and infer $\alpha$,
the between-household infection rate per infectious individual; 
and (3) combine the parameters obtained in Steps (1) and (2), to obtain the reproduction numbers. 
These are described in Sections \ref{step1}, \ref{step2} and \ref{step3} respectively.

In addition, we describe two alternative methods in Sections \ref{indiv} and \ref{growth},
which mainly differ at Step (2).
The method in Section \ref{step2} assumes that only data for the first symptomatic infection in each
household can be reliably obtained;
that is, the data records the day on which each household first has a symptomatic case.
The alternative method in Section \ref{indiv} assumes that 
the day of first symptoms is also available for every individual.
This is discussed after Section \ref{step2}
because infector individuals can be treated as households of size $1$,
making this in most ways a special case of the method discussed in Section \ref{step2}.

Section \ref{growth} describes a simpler method which bypasses Step (2),
at some cost in accuracy, by first assuming an exponential growth rate $r$,
and using that to find the between-household infection rate $\alpha$.

\subsection{Step 1: Infer the within-household parameters}\label{step1}


The first step of the method is to infer within-household parameters from the FFX data.
We use data from within the observed households, and assuming independence between these households,
consistent with the assumption of branching dynamics,
perform inference and hence sample from the joint posterior distribution for the within-household parameters
which were described in Table \ref{tab:modeltab}:
$\beta$, $t_E$, $t_P$, $t_I$ and $p_s$;
though instead of $\beta$ we infer $R_{0i}=\beta(t_P+t_I)$,
the within-household component of the reproduction number $R_0$ or $R_\text{eff}$.

This step is essentially the same as described in our previous paper \cite{Black2017},
using a Markov chain Monte Carlo scheme
to sample from the posterior distribution of the parameters,
so it is not described in detail here. 
The only difference is that instead of an exact matrix exponential approach to calculate the likelihood,
we use a particle-marginal approach to estimate the likelihood \cite{Walker2019},
using importance sampling for the household model \cite{Black2018}.
Despite being an estimate, this still targets the correct posterior distribution \cite{Schon2018}. 
This has the advantage of being able to handle more complex household models and larger household sizes.

The output from this step is $N_S$ samples from the joint posterior of the within-household parameters.
We typically use $N_S=10000$, thinning down from a larger number of posterior
points obtained during the Markov chain Monte Carlo inference.\footnote{While 
  thinning is unnecessary in most applications,
  it can be appropriate when a significant amount of post-processing is performed \cite{Link2012},
  as is the case here.}

Although it is not our direct aim in this paper,
we note that the inference of within-household parameters other than the transmission rate 
provides other valuable information on the epidemic:
the time-related parameters ($t_E$, $t_P$ and $t_I$) provide information 
such as the latent period (time from infection to infectiousness),
incubation period (time from infection to symptoms) and infectious period;
while $p_s$ gives a measure of the severity of the disease;
and this all arises due to the collection of FFX data.
More detailed information on severity can be obtained if extra FFX data is collected,
such as which individuals 
needed to visit a doctor or hospital.


\subsection{Step 2: Infer the between-household infection rate}\label{step2}

The second step of inference is to infer $\alpha$,
the between-household infection rate per infectious individual,
from the daily information of newly infected households.
By assuming an effectively infinite supply of households,
the spread of infection between households can be modelled as a branching process.

To simplify the discussion,
we first consider the case in which all households are the same size
(that is, all households contain the same number of individuals),
and all households contain at least one symptomatic case.
The extensions to different household sizes and households without symptomatic cases,
which add several terms to the equations, but are not difficult conceptually,
are given in Section \ref{diffHHsizes}.

\subsubsection{Fixed household size, all symptomatic}\label{fixed}

For the purposes of between-household inference,
we need only the date of the first symptomatic case in each household.
(Section \ref{indiv} describes the situation in which the dates of all
\emph{individual} symptomatic cases in each household are available.)
Therefore the data 
can be condensed to a single vector of daily observations
$y=(y_1 \dots y_d)$,
where each element $y_j$ is
the number of households which are first symptomatic
(that is, having at least one individual in the household showing symptoms, for the first time) on day $j$.
The inference method is to estimate $\mathbb{E}[y_j]$,
the expected number of newly symptomatic households on day $j$, in terms of $\alpha$,
and hence estimate the posterior distribution of $\alpha$.

In order to estimate $\mathbb{E}[y_j]$,
we need to estimate 
the infection potential of a household:
the sum of the infectious times of all infected individuals in that household. 
Then $\Psi$, the expected value of the infection potential, is given by:
\begin{equation}\label{psi}
  \Psi = \mathbb{E} \left[ \int_0^{\infty} I(t) dt \right] ,
\end{equation}
where $I(t)$ is the number of infectious individuals in a household at time $t$.
This means that each household generates, on average, $\alpha \Psi$ between-household infections.
The value of $\Psi$ can be estimated from simulations of the household model.

\begin{figure}[H]
\begin{tikzpicture}[->,>=stealth',shorten >=1pt,auto,node distance=2.5cm,
    thick,main node/.style={rectangle,text width=1cm,text height=1cm, 
    fill=white,draw,font=\sffamily\Large\bfseries},
    scale=0.5, every node/.style={transform shape}]

  \node[main node, style={text width=5cm, text height=1cm}, draw=white] (infector) at (7, 3) {Infector Household};
  \node[main node, style={text width=5cm, text height=1cm}, draw=white] (infectee) at (7, 0) {Infectee Household};
  \node[main node, style={text width=5cm, text height=1cm}, draw=white] (infectee) at (7, -3) {Dummy\\variable used};

  \node[main node, style={}, draw=white] () at (10, -3) {$l$};
  \node[main node, style={}, draw=white] () at (18, -3) {$k$};
  \node[main node, style={}, draw=white] () at (28, -3) {$j$};

  \node[main node, style={}, draw=white] (l0) at (10, 0) {};
  \node[main node, style={}, draw=white] (j1) at (28, 4) {};

  \node[main node, style={}, draw=gray!80] (l) at (10, 4) {\:E};
  \node[main node, style={}, draw=gray!80] () at (11, 4) {};
  \node[main node, style={}, draw=gray!80] () at (12, 4) {};
  \node[main node, style={}, draw=gray!80] () at (13, 4) {};
  \node[main node, style={}, draw=gray!80] () at (14, 4) {};
  \node[main node, style={}, draw=gray!80] () at (15, 4) {};
  \node[main node, style={}, draw=gray!80] () at (16, 4) {};
  \node[main node, style={}, draw=gray!80] () at (17, 4) {};
  \node[main node, style={}, draw=gray!80] (k) at (18, 4) {FS};
  \node[main node, style={}, draw=gray!80] () at (19, 4) {};
  \node[main node, style={}, draw=gray!80] () at (20, 4) {};
  \node[main node, style={text width=0.8cm}, draw=gray!80] () at (21, 4) {};


  \node[main node, style={}, draw=gray!80] (e) at (20, 0) {\:E}; 
  \node[main node, style={}, draw=gray!80] () at (21, 0) {};
  \node[main node, style={}, draw=gray!80] () at (22, 0) {};
  \node[main node, style={}, draw=gray!80] () at (23, 0) {};
  \node[main node, style={}, draw=gray!80] () at (24, 0) {};
  \node[main node, style={}, draw=gray!80] () at (25, 0) {};
  \node[main node, style={}, draw=gray!80] () at (26, 0) {};
  \node[main node, style={}, draw=gray!80] () at (27, 0) {};
  \node[main node, style={}, draw=gray!80] (j) at (28, 0) {FS};
  \node[main node, style={}, draw=gray!80] () at (29, 0) {};
  \node[main node, style={text width=0.8cm}, draw=gray!80] () at (30, 0) {};

  \path[every node/.style={font=\sffamily\small}]
  ([xshift=-1mm]l.north)  edge [bend left, <->]    node {$A_1$} ([xshift=1mm]k.north)
  ([xshift=-1mm]l0.north)  edge [bend left, <->]    node {$G$} ([xshift=1mm]e.north)
  ([xshift=-1mm]e.north)  edge [bend left, <->]    node {$A_2$} ([xshift=1mm]j.north)
  ([xshift=-1mm]k.north)  edge [bend left, <->, dotted]    node {$S_H$} ([xshift=1mm]j1.north)
  ([xshift=-1mm]l0.south)  edge [bend right=15, <->]    node {$C$} ([xshift=1mm]j.south)
  (l0)  edge [style=dashed, -] node {}  (l.south)
  (j)  edge [style=dashed, -] node {}  (j1.north)

  
  ;
\end{tikzpicture}
\caption{Schematic representation of the random variables $A_1$, $A_2$, $G$, $S_H$ and $C$.
  Time progresses from left to right, and each square represents a day.
  ``E'' denotes the day a household is first exposed and infected;
  and ``FS'' denotes the first day of symptoms in a household.
  A dotted line is used to show $S_H$, the household-wise equivalent of the serial interval,
  because it is discussed but not used.
  The bottom row shows the dummy variables used for each day in the analyses in Sections \ref{step2} and \ref{indiv}.
  }
\label{ABC}
\end{figure}
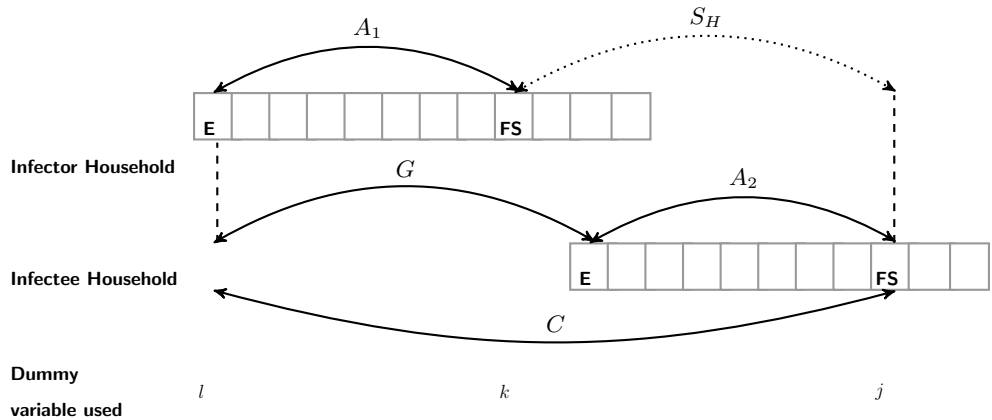

We also need the distributions of $A_1$, $A_2$ and $G$,
as shown in Figure \ref{ABC}.
That is, for any given pair of infector and infectee households,
we define:
$A_1$ to be a random variable denoting the time from infection to first symptoms in the infector household
(the household-wise equivalent of the incubation period  \cite{CDC2012}),
rounded to the nearest day;
$A_2$ to be a random variable denoting the time from infection to first symptoms in the infectee household,
rounded to the nearest day;
and $G$ to be a random variable denoting the time from first infection in the infector household to first infection in the infectee household
(the household-wise equivalent of the generation time \cite{Britton2019}),
rounded to the nearest day.
We also define $S_H$ to be a random variable denoting
the time from first symptoms in the infector household to first symptoms in the infectee household
(the household-wise equivalent of the serial interval \cite{Gostic2020});
and $C$ to be the sum of $G$ and $A_2$, 
making it the time from first infection of the infector household,
to first symptoms in the infectee household, rounded to the nearest day.

An important part of our method is that, for a given sample of the within-household parameters,
we can obtain unbiased estimates for the distributions of $A_1$, $A_2$ and $G$ (and hence $C$)
from simulation of the within-household model, without the need for contact tracing.

The analysis would be simpler if we could use $S_H$,
the household-wise equivalent of the serial interval,
to estimate $\mathbb{E}[y_j]$.
However
an unbiased distribution of $S_H$ cannot be obtained without accounting for
the growth or decay in the outbreak \cite{Champredon2015},
in terms of the number of newly infected households per day. 
This is because a sample of $S_H$ is obtained by subtracting a sample of $A_1$ from a sample of $C$;
and this subtraction of $A_1$, also referred to as looking backwards in time \cite{Britton2019},
biases the distribution depending on whether the outbreak is growing or decaying,
as has been pointed out previously
\cite{ScaliaTomba2010, Britton2019, Champredon2015}. 

However it is possible to estimate $\mathbb{E}[y_j]$ without $S_H$ and without contact tracing,
using the other variables in Figure \ref{ABC},
because these are estimated forward from the day the infector household is exposed
(day $l$ in Figure \ref{ABC}).
To do this, it is useful to define $w_{(j,k,l)}$,
the expected number of between-household infections
which are caused by an infector household which was infected on day $l$
and first symptomatic on day $k$,
and which are first symptomatic in an infectee household on day $j$.
Then consideration of each day $j$ gives:
\begin{equation}\label{w_j}
  \mathbb{E}[y_j] = \sum_k \sum_l w_{(j,k,l)} . 
\end{equation}
Since every infector household generates on average $\alpha \Psi$ between-household infections,
it follows from consideration of each day $k$:
\begin{equation}\label{w_k}
\alpha \Psi  \mathbb{E}[y_k] = \sum_{j} \sum_{l} w_{(j,k,l)} ;
\end{equation}
and considering infections generated from households initially infected on day $l$ gives:
\begin{equation}\label{w_l}
w_{(j,k,l)} = \alpha \Psi T_l \,p(A_1=k-l, C=j-l) ,
\end{equation}
where $T_l$ is the true number of households infected on day $l$.
We cannot measure $T_l$, but
a reasonable approximation is that it is proportional to $y_l$,
as this would be the case
if the rate of growth or decay was constant
and if there was no stochasticity.
So we estimate $T_l$ to be $y_l$ multiplied by some value $K_k$ which is constant for each value of
$k$, giving:
\begin{equation}\label{w_l2}
  w_{(j,k,l)} = \alpha \Psi y_l K_k \,p(A_1=k-l, C=j-l) .
\end{equation}
We estimate $K_k$ by substituting \eqref{w_l2} into \eqref{w_k} and replacing $\mathbb{E}[y_k]$
with the observed value $y_k$, giving:
\begin{equation*}
\alpha \Psi y_k = \sum_{j} \sum_{l} \alpha \Psi y_{l} K_k \,p(A_1=k-l, C=j-l) ,
\end{equation*}
\begin{equation}\label{K}
\Rightarrow K_k = \frac{y_k}{\sum_{j} \sum_{l} y_{l} \,p(A_1=k-l, C=j-l)} .
\end{equation}
Since all days $l$ will be the same as or earlier than day $k$, it follows that
$K_k > 1$ when the number of new infections per day is growing at day $k$,
$K_k < 1$ when it is falling,
and $K_k \approx 1$ when it is approximately constant.
So the $K_k$ factor provides a correction to account for the growth or decay of the outbreak.
Appendix \ref{Keffect} illustrates how inference results can be
biased if this term is not included.

Then substituting \eqref{w_l2} into \eqref{w_j} gives:
\begin{equation*}
  \mathbb{E}[y_j] = \alpha \Psi \sum_k \sum_l y_l K_k \,p(A_1=k-l, C=j-l) .
\end{equation*}


As shorthand, we define the quantity $\xi_j$ which is equal to $\mathbb{E}[y_j] / \alpha$, that is:
\begin{equation}\label{xi}
  \xi_j = \Psi \sum_k \sum_l y_l K_k \,p(A_1=k-l, C=j-l) .
\end{equation}

We use a gamma distribution as the prior of $\alpha$, 
meaning that if the data is Poisson distributed,
then the posterior distribution of $\alpha$ is also a gamma distribution \cite{Cori2013, Thompson2019}.
Using $c$ days of data up to and including day
$d$, 
it can be shown (Appendix \ref{conjproof}) that the posterior of $\alpha$ is
\begin{equation}\label{post1}
  f(\alpha|y) \sim
  \text{Gamma}\left(a+\sum_{j=d+1-c}^d y_j, \quad b+\sum_{j=d+1-c}^d \xi_j \right) ,
\end{equation}
where the first parameter is the shape and the second parameter is the rate,
and $a$ and $b$ are the prior's shape and rate respectively.
This equation gives the distribution of $\alpha$ as calculated on day $d$.

Aside from the observed data values in $y$,
all quantities on the right hand sides of \eqref{K} and \eqref{xi}
can be estimated from simulations of the household model.
Thus this step
gives 
samples from the joint posterior distribution of $\alpha$ and the within-household parameters.

By making $K_k$ constant for any value of $k$,
we are assuming a constant rate of growth or decay.
A more sophisticated calculation may be possible which 
accounts for change in this rate.
But since we are only using the last $c$ days of data for inference 
in \eqref{post1},
this should not be necessary. 
However, it may be a point for investigation in the future.

\subsubsection{General case}\label{diffHHsizes}

We modify the discussion in the previous section to account for households of different sizes,
and asymptomatic cases.

To account for asymptomatic cases,
we define $p_h$ to be the probability of an infected household being symptomatic, and
$p_i$ to be the probability of a randomly chosen infectious individual being a member of a symptomatic household.
We define a symptomatic household
to be an infected household in which at least one individual is eventually symptomatic;
and an asymptomatic household
to be an infected household in which all infected individuals remain asymptomatic.

To account for different household sizes,
we require the observed data to be a 2-dimensional array $y$,
with $y_{(j,n)}$ being the number of households of size $n$ which are first symptomatic on day $j$.
We replace $\Psi$ with $\Psi_m$, the expected infection potential of a symptomatic household of size $m$, given by:
\begin{equation}\label{psi_m}
\Psi_m = \mathbb{E} \left[ \int_0^{\infty} I_m(t) dt \right] ,
\end{equation}
where $I_m(t)$ is the number of infectious individuals in a symptomatic household of size $m$ at time $t$.
In addition, $M$ and $N$ are random variables denoting
the sizes of the infector and infectee households respectively.

We define the quantity $w_{(j,k,l,m,n)}$,
which is the expected number of between-household infections
which are caused by an infector household of size $m$ which was infected on day $l$
and first symptomatic on day $k$,
and which are first symptomatic in an infectee household of size $n$ on day $j$.

Then consideration of each day $j$ also needs to account for the fact that on average
only $p_i$ of all infections come from infectious households,
and only $p_h$ of infections are in infectee households which become symptomatic.
So \eqref{w_j}, \eqref{w_k} and \eqref{w_l} are modified to:
\begin{equation}\label{w_jm}
  \mathbb{E}[y_{(j,n)}] = \frac{p_h}{p_i} \sum_k \sum_l \sum_m w_{(j,k,l,m,n)} ;
\end{equation}
\begin{equation}\label{w_km}
\alpha \Psi_m  \mathbb{E}[y_{(k,m)}] = \sum_{j} \sum_{l} \sum_{n} w_{(j,k,l,m,n)} ; \text{ and}
\end{equation}
\begin{equation}\label{w_lm}
w_{(j,k,l,m,n)} = \alpha \Psi_m T_{l,m} \,p^*(A_1=k-l, C=j-l, N=n|M=m) ;
\end{equation}
where $T_{(l,m)}$ is the true number of symptomatic households of size $m$ infected on day $l$,
and $p^*$ means \emph{``the probability, given that both the infector and infectee households are symptomatic''}.

Then following a similar analysis to Section \ref{fixed} gives:
\begin{equation}\label{Km}
  K_{(k,m)} = \frac{y_{(k,m)}}
  {y_{(l,m)} \,p^*(A_1=k-l, C=j-l, N=n|M=m)} 
\end{equation}
and
%
%
\begin{equation}\label{xim}
  \xi_{(j,n)} = \frac{p_h}{p_i} \sum_k \sum_l \sum_m \Psi_m y_{(l,m)} K_{(k,m)} \,p^*(A_1=k-l, C=j-l, N=n|M=m) ,
\end{equation}
where $\xi_{(j,n)} = \mathbb{E}[y_{(j,n)}] / \alpha$.

We again use a gamma distribution as the prior of $\alpha$, 
and infer using $c$ days of data up to and including day $d$.
This means \eqref{post1} is modified to:
\begin{equation}\label{post1m}
  f(\alpha|y) \sim
  \text{Gamma}\left(a+\sum_{j=d+1-c}^d \sum_n y_{(j,n)},
  \quad b+\sum_{j=d+1-c}^d \sum_n \xi_{(j,n)} \right) ,
\end{equation}
where the first parameter is the shape and the second parameter is the rate,
and $a$ and $b$ are the prior's shape and rate respectively.

  Once again,
aside from the observed data values in $y$,
all quantities on the right hand sides of \eqref{Km} and \eqref{xim}
can be estimated from simulations of the household model.
Therefore, for each of the $N_S$ samples from the joint posterior of the within-household parameters obtained in Step 1,
we run a number of simulations of the household model (typically 1000) per household size,
allowing $\xi_{(j,n)}$ to be estimated for all values of $n$, for each day $j$;
and then a single sample of $\alpha$ is taken from \eqref{post1m}.
Thus this step
gives $N_S$ samples from the joint posterior distribution of $\alpha$ and the within-household parameters.

\subsection{Step 3: Calculating reproduction numbers and growth rate}\label{step3}

Once samples of all the model parameters are obtained,
the population reproduction numbers are fairly straightforward to evaluate.

To account for different household sizes,
we define $h_m$ to be the probability that a randomly chosen household will be of size $m$.
Therefore a newly infected household has size $m$ with probability $\pi_m$,
where $\pi_m$ is the size-biased distribution \cite{Black2013},
and is given by the equation,
\begin{equation}\label{pi_m}
  \pi_m = \frac{m h_m}{\sum_i i h_i} .
\end{equation}

The household reproduction number $R_*$ \cite{Black2013}
can then be evaluated using previously obtained values:
\begin{equation}\label{Rstar}
  R_* =  \alpha \sum_m \pi_m \Psi_m ,
\end{equation}
where $\Psi_m$ is given in \eqref{psi_m}.


We use the definition of $R_0$ as the expected number of infections
directly caused by the very first infectious individual in the population.
For each value of the household size $m$,
one can run simulations of the household model and estimate $g_m$,
the mean number of infections caused directly by the first infected member of the household, giving:
\begin{equation}\label{R0pure}
R_0 =  \sum_m \pi_m g_m .
\end{equation}
This equation can also be used to calculate $R_\text{eff}$.

Other definitions of $R_0$ and $R_\text{eff}$ have been proposed for two-level models \cite{Ball1997,Goldstein2009,Pellis2012}.
These can also be calculated using the raw data from the previous steps.
For instance, an alternative might be to use
$R_{HI}$ \cite{Goldstein2009},
the expected number infected by a randomly chosen infectious individual.
In that case, we estimate $f_m$, the average number of infections within an infected household of size $m$,
from simulations of the household model.
Between them, the $f_m$ household members generate on average
$(f_m-1)$ within-household infections and $\alpha \Psi_m$ between-household infections, giving:
\begin{equation}\label{RHI}
R_{HI} =  \sum_m \pi_m \frac{f_m-1 + \alpha \Psi_m}{f_m} .
\end{equation}

An estimate of the exponential growth rate $r$ can also be obtained \cite{Black2013}, by finding the value of $r$ for which
\begin{equation}\label{findr}
   \alpha \sum_m \pi_m \mathbb{E} \left[ \int_0^\infty I_m(t) e^{-rt} dt \right] = 1 .
\end{equation}

Equation \eqref{findr} includes a path integral, and
its expectation can be calculated by solving a set of linear equations \cite{Pollett2002,Pollett2003,Ross2010},
which are written in terms of the generator of the within-household CTMC.
However this has a practical drawback: the calculations each require a square matrix with one row and one column
for each non-absorbing state of the household model,
and so the size of such matrices increases exponentially with the household size $m$.
In our experience these matrices get infeasibly large,
in terms of computer memory and execution time,
for $m$ over $8$ or so household members.

A simpler and more practical approach is to also estimate this quantity with the aid of simulation.
Any simulation of $I_m(t)$ is a piecewise step function, 
meaning $\int_0^\infty I_m(t) e^{-rt} dt$, and hence the left hand side of \eqref{findr},
can be efficiently estimated by simulation for any given value of $r$;
allowing $r$ to be found 
using a root finding program \cite{Juliaroots}.
The simulations are not memory-intensive and so do not have the same limitations
as the system of linear equations method.
Another advantage is that, by incorporating the summation into an existing simulation model,
the code is relatively easy to verify.

In all cases, these parameters are evaluated as samples from the distribution of the posterior.
The output from Step (2) is $N_S$ samples from the joint posterior of $\alpha$ and the within-household parameters;
we typically use $N_S=10000$.
The method in Step (3),
to evaluate any of $R_*$, $R_0$, $R_\text{eff}$ or $r$,
is repeated for each of these samples.
Therefore the method gives $N_S$ joint samples of all parameters.


\subsection{Observed individuals}\label{indiv}

In this section and  in Section \ref{growth},
we present two alternate methods,
which are relevant if different data is available.

The analysis in Section \ref{step2} assumes that the only observed data
is the number of newly symptomatic households of each household size on each day,
which is stored in the 2-dimensional array $y$.
In this section we consider the situation in which
the number of newly symptomatic \emph{individuals} is also available.
We denote this as $z = (z_1 \dots z_d)$,
where $z_j$ is the number of individuals who are first symptomatic on day $j$.

In that case, there is no need to use the infector households to estimate the
number of newly infectious individuals, because that quantity is
available directly as $z$.
The two-dimensional array $y$ is still used as the count of new infectee households.

Infector individuals can be treated identically to infector households of size $1$,
which are symptomatic with probability $p_s$ (Table \ref{tab:modeltab}).
So we modify the analysis of Section \ref{diffHHsizes}
by replacing $y_{(l,m)}$ with $z_l$ for all $l$,
replacing $y_{(k,m)}$ with $z_k$ for all $k$, 
setting $m=1$ as the only possible value of $M$,
and replacing $p_i$ with $p_s$.
This means \eqref{Km} is modified to:
\begin{equation}\label{Km_indiv}
K_{(k,1)} = \frac{z_k}{\sum_{j} \sum_{l} \sum_{n} z_{l} \,p^*(A_1=k-l, C=j-l, N=n|M=1)} ;
\end{equation}
and \eqref{xim} is modified to:
\begin{equation}\label{xim_indiv}
  \xi_{(j,n)} = \frac{p_h}{p_s} \sum_k \sum_l \Psi_1 z_{l} K_{(k,1)} \,p^*(A_1=k-l,C=j-l, N=n|M=1) .
\end{equation}

The rest of the analysis proceeds as in Sections \ref{diffHHsizes} and \ref{step3},
including the use of \eqref{post1m} to find the distribution of $\alpha$.

\subsection{Inferring $r$ directly, and $\alpha$ from $r$}\label{growth}

Another alternative method, which builds on a previously published method \cite{Ross2010, Black2013},
is to first infer the growth rate $r$,
and then use $r$ to find $\alpha$.
This estimate of $\alpha$, along with the estimates of the within-household parameters obtained in Step 1,
is then used to obtain estimates of $R_*$, $R_0$ or $R_\text{eff}$.
The inference of $r$ is an extra step in this inference, so this method is usually not preferred.
However it might be preferable in situations where the exponential growth rate is relatively easy to obtain.
For instance, the distribution of $r$ could be approximately inferred by performing a Bayesian linear regression
of log$(y)$ versus time,
where $y$ is the number of newly infected households per day.

Given samples from the posterior distribution of $r$ obtained by this or another method,
we can take samples from the posterior of the within-household parameters,
and calculate corresponding samples of the between-household infection rate $\alpha$
using \eqref{findr}, but with $\alpha$ rather than $r$ being the unknown quantity.
As in Section \ref{step3}, usually the simplest and most efficient way to find $\alpha$
is to estimate the integral in \eqref{findr} by simulation.

With samples of $\alpha$ generated, 
corresponding samples of $R_*$, $R_0$ or $R_\text{eff}$ can then be generated
using \eqref{Rstar}, \eqref{R0pure} or \eqref{RHI} as appropriate.

This method has the appeal of bypassing the Section \ref{step2} calculation of $\alpha$.
However a disadvantage is that the samples of $r$ are not joint
with the samples of the within-household parameters.
This is likely to contribute to an overestimation in the variances
in the distributions of $\alpha$, $R_*$, $R_0$ and $R_\text{eff}$.


\section{Testing and Results}\label{impl}



\begin{figure}[H]
\begin{subfigure}{.9\textwidth}
  \centering
  \includegraphics[width=1\linewidth]{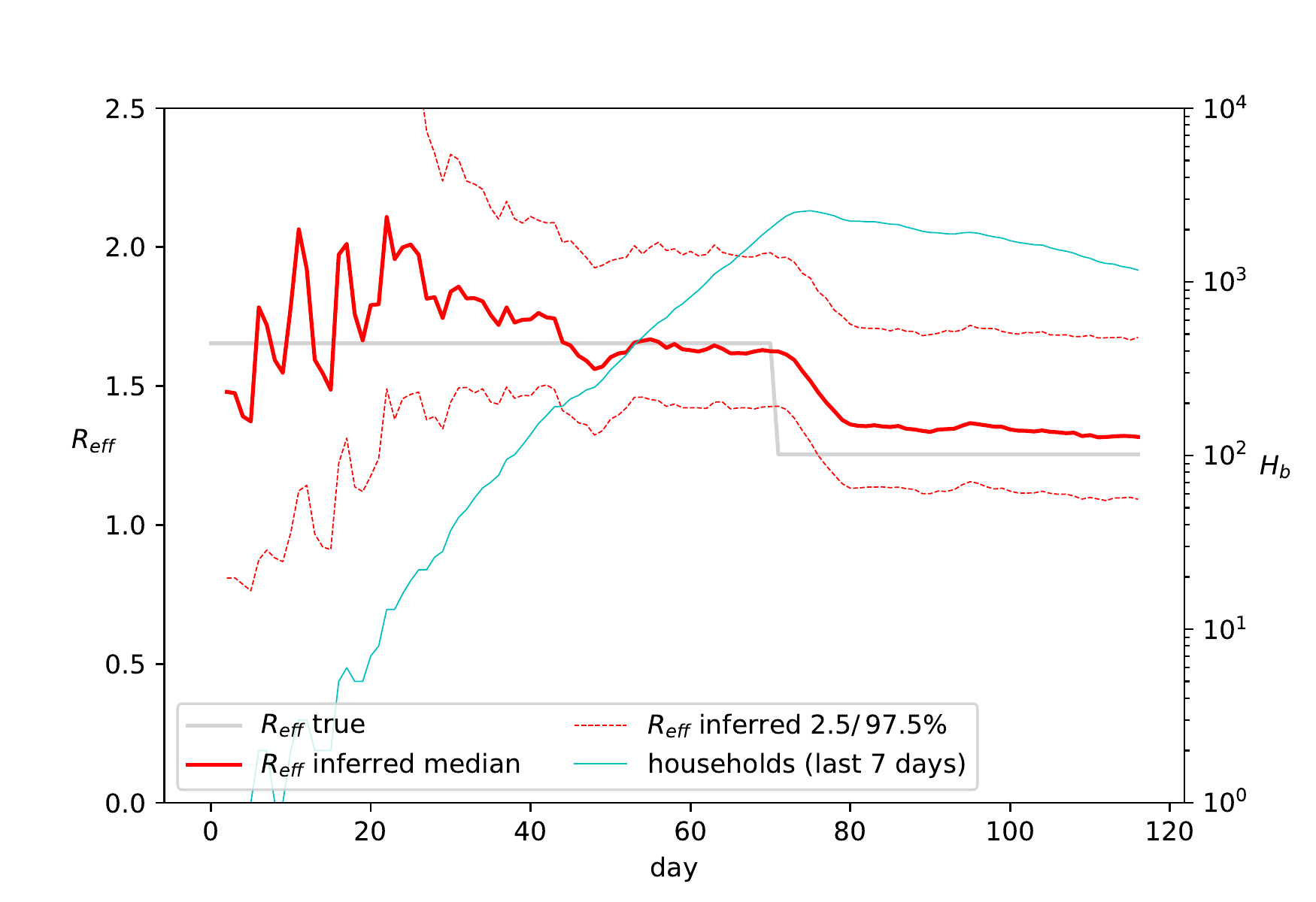}
  \caption{} 
\end{subfigure}
\begin{subfigure}{.9\textwidth}
  \centering
  \includegraphics[width=1\linewidth]{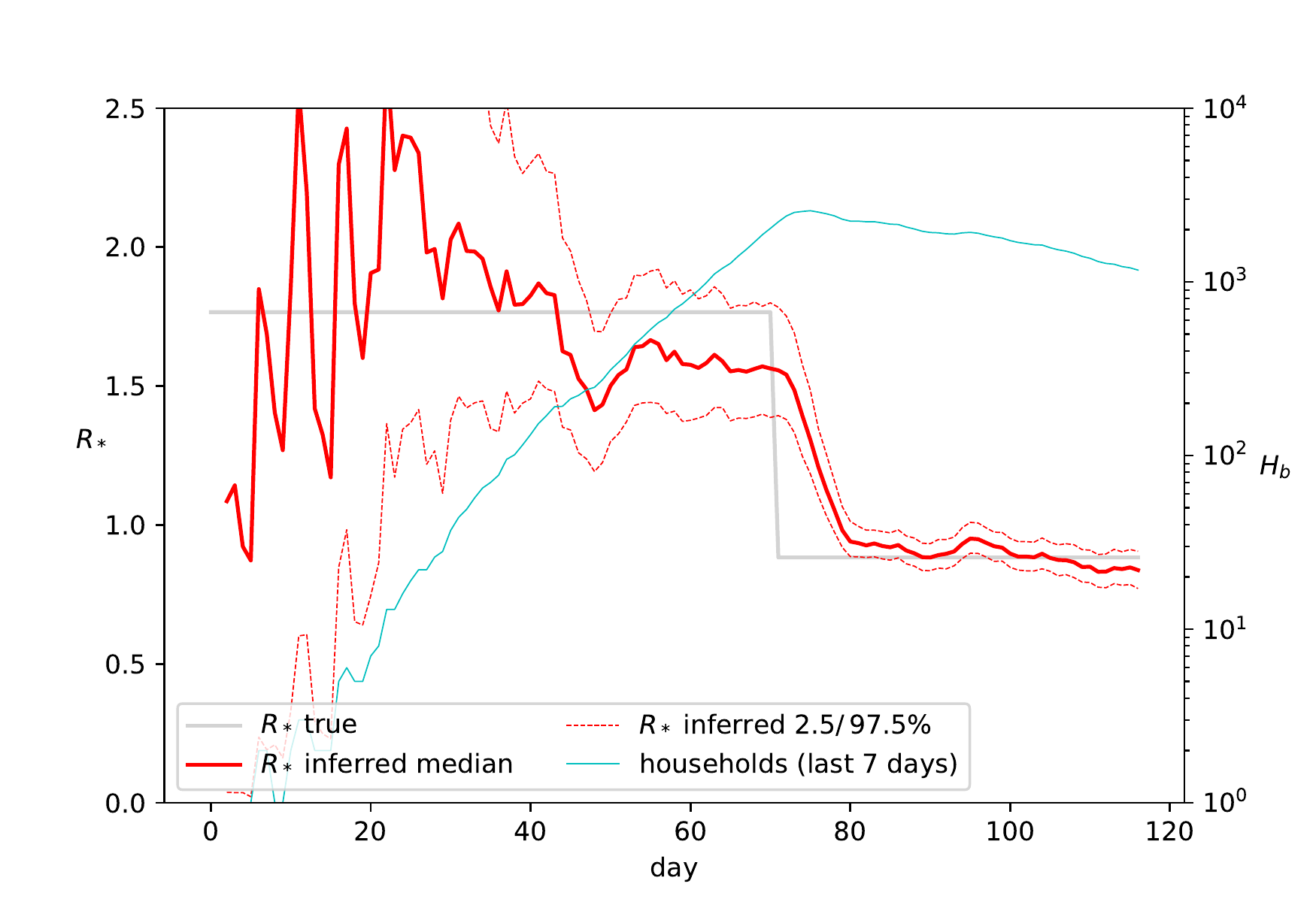}
  \caption{} 
\end{subfigure}
\caption{Inference of (a) $R_\text{eff}$ and (b) $R_*$
  during a simulated outbreak:
  between-household inference updated daily, using the most recent $7$ days of data;
  within-household parameters inferred from 50 households' FFX data.
  The dotted red lines show the $2.5$ and $97.5$ percentiles of the posterior;
  the true value of $R_\text{eff}$ or $R_*$ is shown in grey;
  and the number of households infected in the most recent 7 days ($H_b$) is shown in light blue.
  }
  \label{time_50}
\end{figure}
\begin{figure}[H]
\begin{subfigure}{.9\textwidth}
  \centering
  \includegraphics[width=1\linewidth]{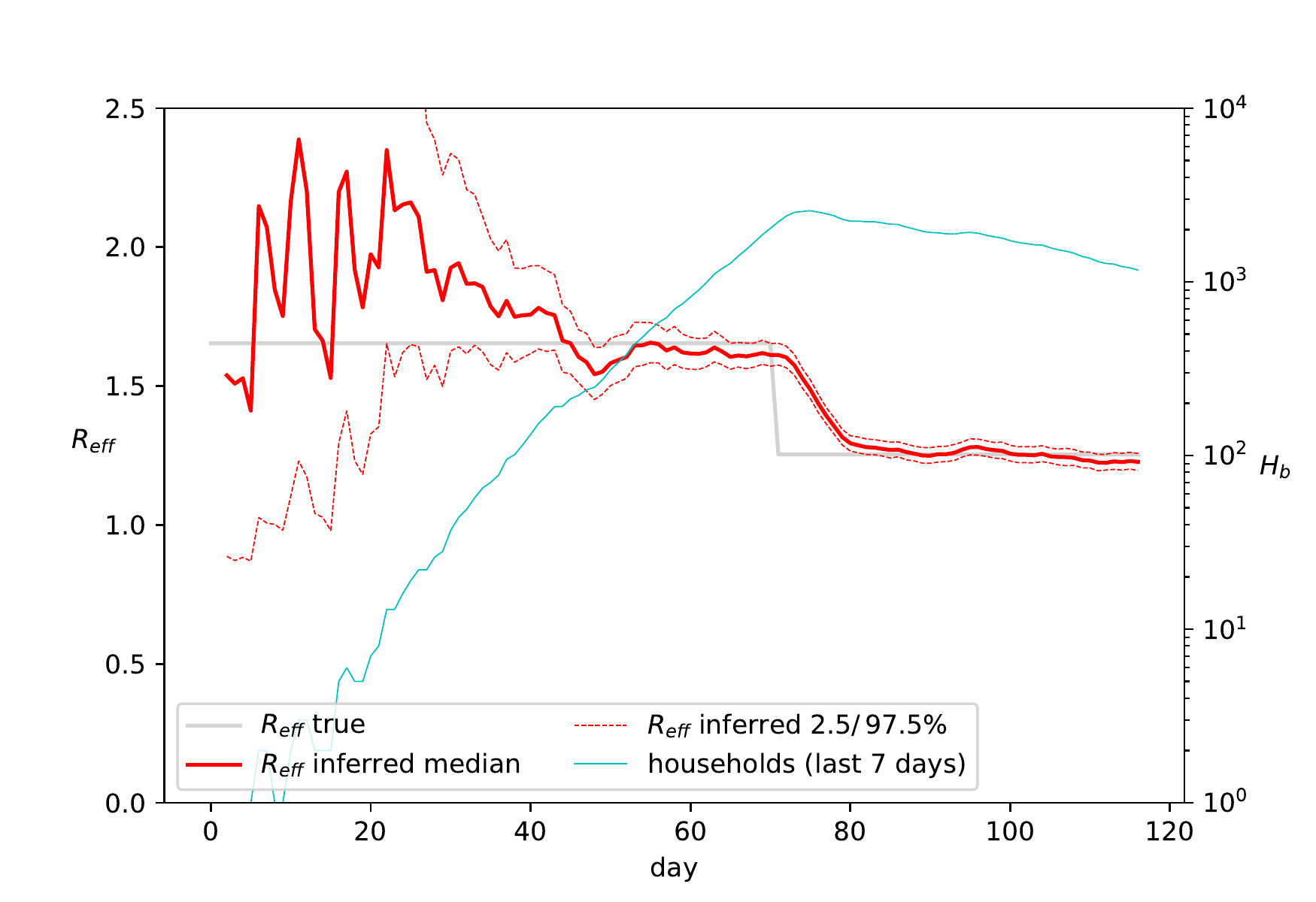}
  \caption{} 
\end{subfigure}
\begin{subfigure}{.9\textwidth}
  \centering
  \includegraphics[width=1\linewidth]{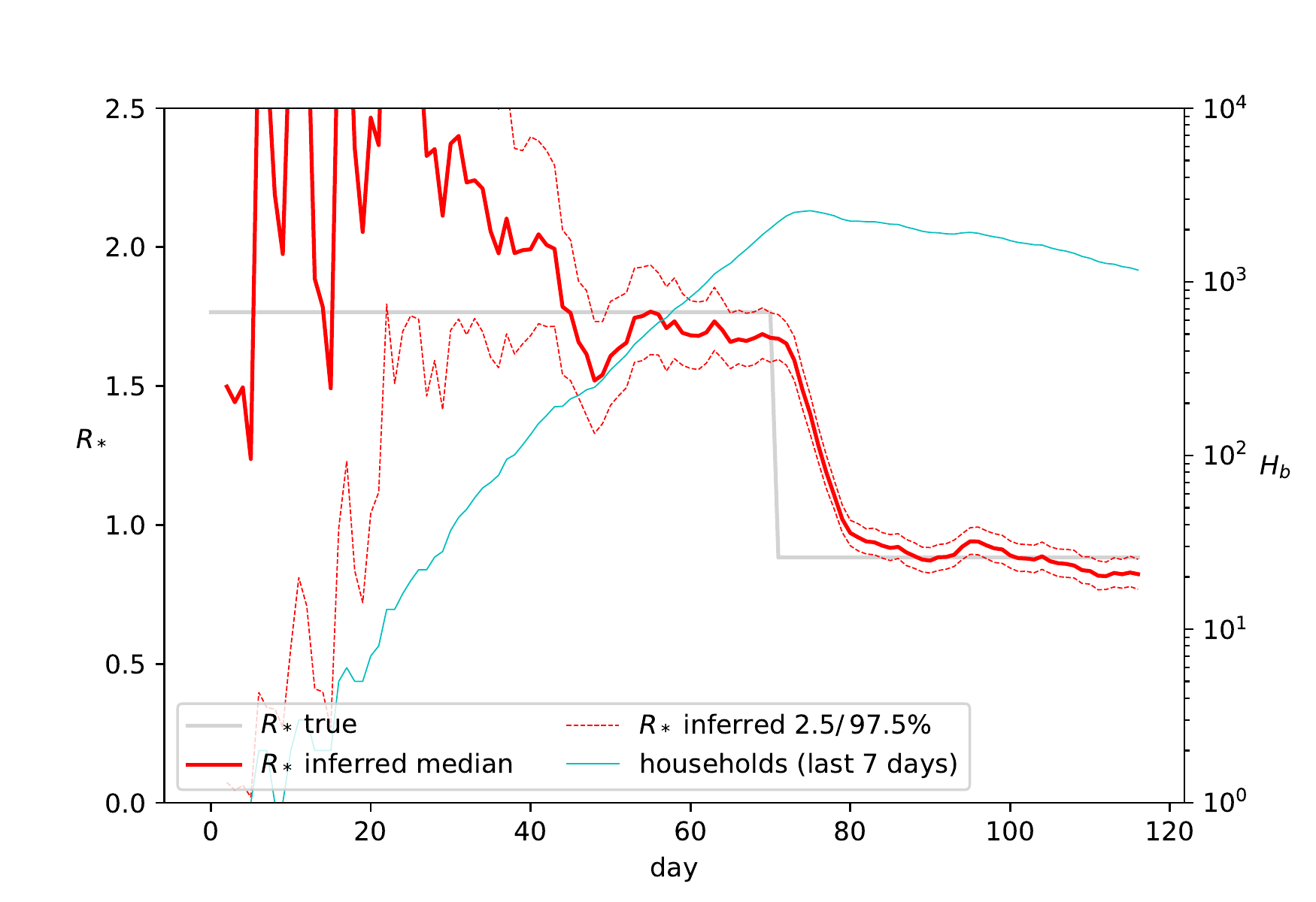}
  \caption{}
\end{subfigure}
\caption{Inference of (a) $R_\text{eff}$ and (b) $R_*$
  during a simulated outbreak:
  between-household inference updated daily, using the most recent $7$ days of data;
  and the within-household parameters known exactly.
  The dotted red lines show the $2.5$ and $97.5$ percentiles of the posterior;
  the true value of $R_\text{eff}$ or $R_*$ is shown in grey;
  and the number of households infected in the most recent 7 days ($H_b$) is shown in light blue.
  }
  \label{time_fixed}
\end{figure}

For the first two sets of tests, in Figure \ref{time_50} and \ref{time_fixed},
data was 
generated using the parameters: 
$t_E=2$, $t_P=1.8$, $t_I=1.5$, $p_s=0.8$,
$R_{0i}=\beta(t_P+t_I)=1.4$;
and $\alpha$ initially $0.242$, then reducing to $0.121$ on day $70$. 
%
For $h_m$, 
the proportions of households of size $m$, and hence the values of the
size-biased distribution $\pi_m$ (Section \ref{step3}),
we used household size data from the 2016 Australian Census \cite{censusHHweb}, capped at a size of 6,
as shown in Table \ref{tab:pop}.
\begingroup
\renewcommand{\arraystretch}{1.5}
\begin{table}[H]
\begin{tabular}{lcll}
\hline
$m$ & number of households & $h_m$ & $\pi_m$ \\
\hline
1 & 2023537 & 0.244 & 0.095 \\
2 & 2768286 & 0.334 & 0.260 \\
3 & 1338376 & 0.162 & 0.188 \\
4 & 1313551 & 0.159 & 0.246 \\
5 &  557262 & 0.067 & 0.131 \\
6 &  284067 & 0.034 & 0.080 \\
\hline
\end{tabular}
\caption{Household size data, from the 2016 Australian Census \cite{censusHHweb}, capped at a size of 6.
  $h_m$ is the proportion of households which are of size $m$,
  while $\pi_m$ is the proportion of individuals who live in a household of size $m$.
  These values of $h_m$ and $\pi_m$ are used throughout the tests in this section.
}
\label{tab:pop}
\end{table}
\endgroup

Inference was performed using the method described in Sections \ref{step1} to \ref{step3},
using \eqref{R0pure} to evaluate $R_\text{eff}$ and \eqref{Rstar} to evaluate $R_*$. 
(All tests were repeated using the method in described in Section \ref{indiv}, with very similar results).
The following priors were used for the within-household parameters:
$R_{0i}$ had a gamma distribution with shape $3$ and rate $5/3$ (mean = $1.8$);
$t_E$ had a gamma distribution with shape $3$ and rate $0.5$ (mean=$6$);
$t_P$ and $t_I$ both had a gamma distribution with shape $3$ and rate $2$
(mean=$1.5$); 
and the prior for $p_s$ had a beta distribution with $\alpha=2.8$ and $\beta=1.2$ (mean=$0.7$).
For the between-household infection parameter $\alpha$,
we followed previous work \cite{Thompson2019} and chose a fairly broad prior:
a gamma distribution with shape $1$ and scale $0.3$ (mean=$0.3$).
With a prior mean infectious time ($t_P+t_I$) of $3$,
  this corresponds to a mean of $0.9$ for the between-household component of $R_\text{eff}$.

For the first set of tests,
we ran the within-household inference only once,
inferring the within-household parameters from the full FFX data for the first $50$ households,
but used daily updates of data for the between-household inference.
This was partly a decision of practicality:
the between-household inference is far faster to run than
the within-household inference, 
due to the former's use of conjugate priors.
It also reflects the authors' experience of FFX data collection in Australia during the COVID-19 outbreak,
in which updates of within-household data tended to occur  slowly.
But it is also based on a reasonable assumption:
that as countermeasures such as social distancing and working from home
are brought in to restrict an outbreak,
primarily the between-household infection rate will be affected,
and the within-household parameters will be impacted less.

We performed inference of $R_\text{eff}$ and $R_*$ for each day of the simulated outbreak,
using the most recent $7$ days of data for the between-household inference,
and the results are shown in Figure \ref{time_50}.
It shows that inference works well, even with only $50$ households of FFX data for within-household inference:
it tracks changes to the transmission rate,
and the confidence interval reduces as the amount of data grows.

In order to determine whether the within-household or between-household inference
was the main source of variance,
the inference was repeated with the within-household parameter
values perfectly known (that is, every sample uses the within-household parameters values used to generate the data:
$R_{0i}=1.4$, $t_E=2$, $t_P=1.8$, $t_I=1.5$ and $p_s=0.8$),
with the results in Figure \ref{time_fixed}.

This shows a substantial reduction in variance,
although this is much more pronounced in $R_\text{eff}$ than $R_*$.
This is because $R_\text{eff}$, by its very definition in \eqref{R0pure}
has a substantial within-household component: within-household and between-household infections are counted.
In contrast, the evaluation of $R_*$ only directly considers between-household transmission,
so changes to the within-household inference method
have a less dramatic effect on $R_*$.

\begin{figure}[H]
\begin{subfigure}{.8\textwidth}
  \includegraphics[width=1\linewidth]{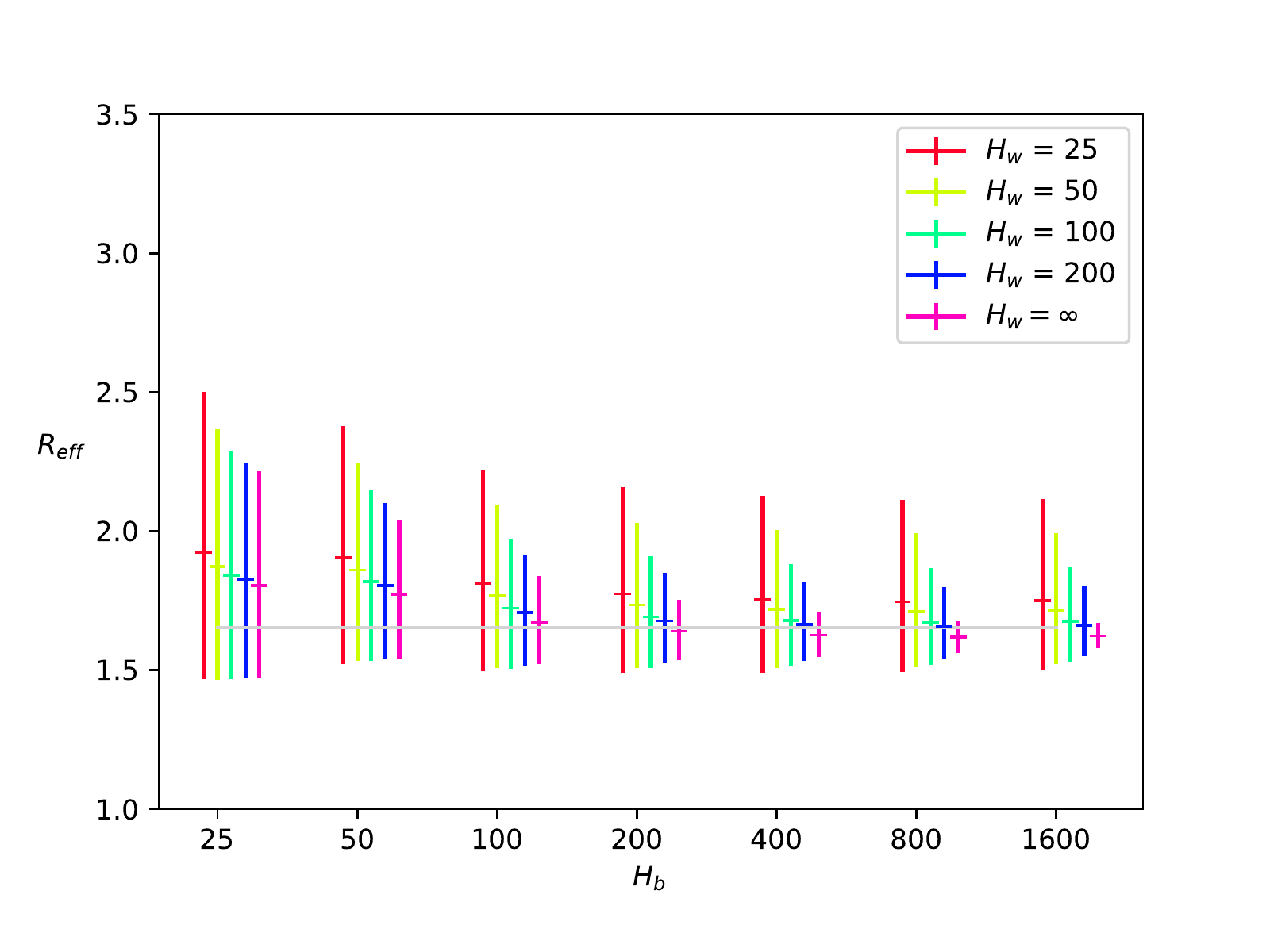}
  \vspace{-1.5\baselineskip}
\caption{}
\end{subfigure}
\begin{subfigure}{.8\textwidth}
  \includegraphics[width=1\linewidth]{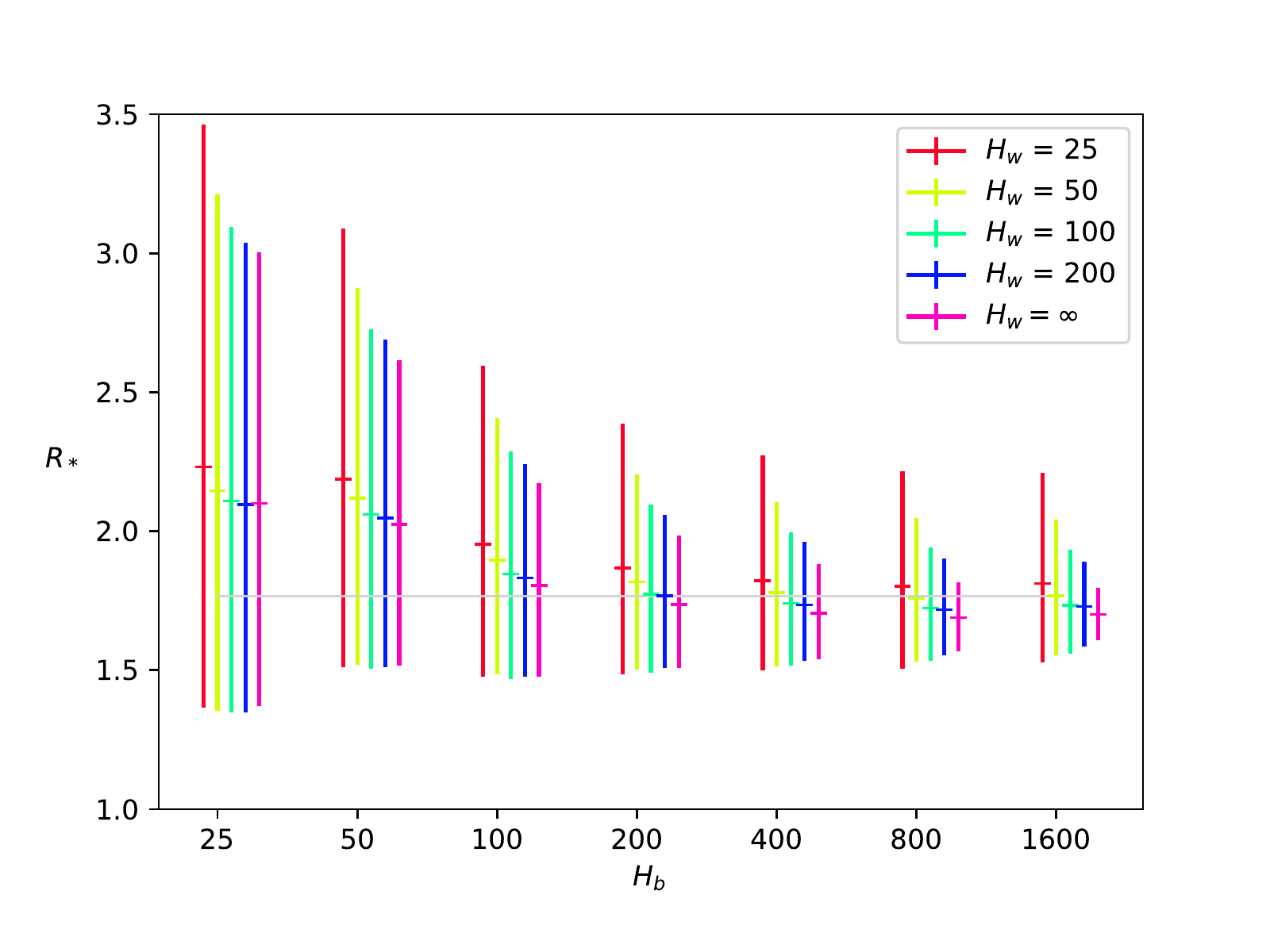}
  \vspace{-1.5\baselineskip}
\caption{}
\end{subfigure}
\caption{
  Plots illustrating the mean and distribution of the posterior of
  (a) $R_\text{eff}$ and (b) $R_*$,
  averaged over the inference results of 40 realisations,
  for different values of $H_w$ (numbers of households used for within-household inference,
  with $H_w=\infty$ meaning the within-household parameters are known exactly)
  and $H_b$ (numbers of households used for between-household inference).
  The horizontal lines show the means of the 40 posterior distributions,
  while the lower and upper extremities show the
  means of the posterior distributions' $2.5$ percentiles,
  and the means of the posterior distributions' $97.5$ percentiles, respectively.
  The true values ($1.654$ and $1.766$ respectively) are shown as grey lines.
}
\label{run5a}
\end{figure}

  We tested the method further by aggregating the results from 40 different realisations,
  across a range of values of
  $H_w$ (the number of households used for within-household inference)
  and $H_b$ (the number of households used for between-household inference),
  with the results shown in Figure \ref{run5a}.
Each of the 40 realisations was 
generated using the parameters:
$R_{0i}=1.4$, $t_E=2$, $t_P=1.8$, $t_I=1.5$, $p_s=0.8$, $\alpha=0.242$
and the distribution of household sizes in Table \ref{tab:pop};
corresponding to $R_\text{eff}=1.654$ and $R_*=1.766$.
As in Figures \ref{time_50} and \ref{time_fixed},
the parameter $H_b$ refers to the number of houses newly infected in the most recent 7 days.
For within-household inference, we also included the results
assuming the within-household parameters are known exactly,
effectively corresponding to $H_w=\infty$.

Overall, these results show that the inference method works well.
We see that accuracy improves, and variance decreases, as more data becomes
available; and that this is the case for both within-household and between-household inference.
This is similar to what is observed in Figures \ref{time_50}  and \ref{time_fixed}.
We also see that one or other component can dominate the variance.
That is, if $H_w$ is small, there is not a lot to be gained by increasing $H_b$;
and vice versa.

We do notice that both $R_\text{eff}$ and $R_*$ are overestimated
when there is a small amount of data, in particular
for lower values of $H_b$.
In the case of lower values of $H_b$, 
we believe that the main cause is that
the initial fade-out of some realisations causes a ``selection bias'', leading
to an overestimation of $R_\text{eff}$ and $R_*$ \cite{Mercer2011, Rebuli2018},
which is not simple to correct.

\subsection{Forecasting}

The inferred values of the exponential growth rate $r$ can also be used
for simple forecasting, assuming a continued exponential growth or
decay.

\begin{figure}[H]
  \includegraphics[width=1\linewidth]{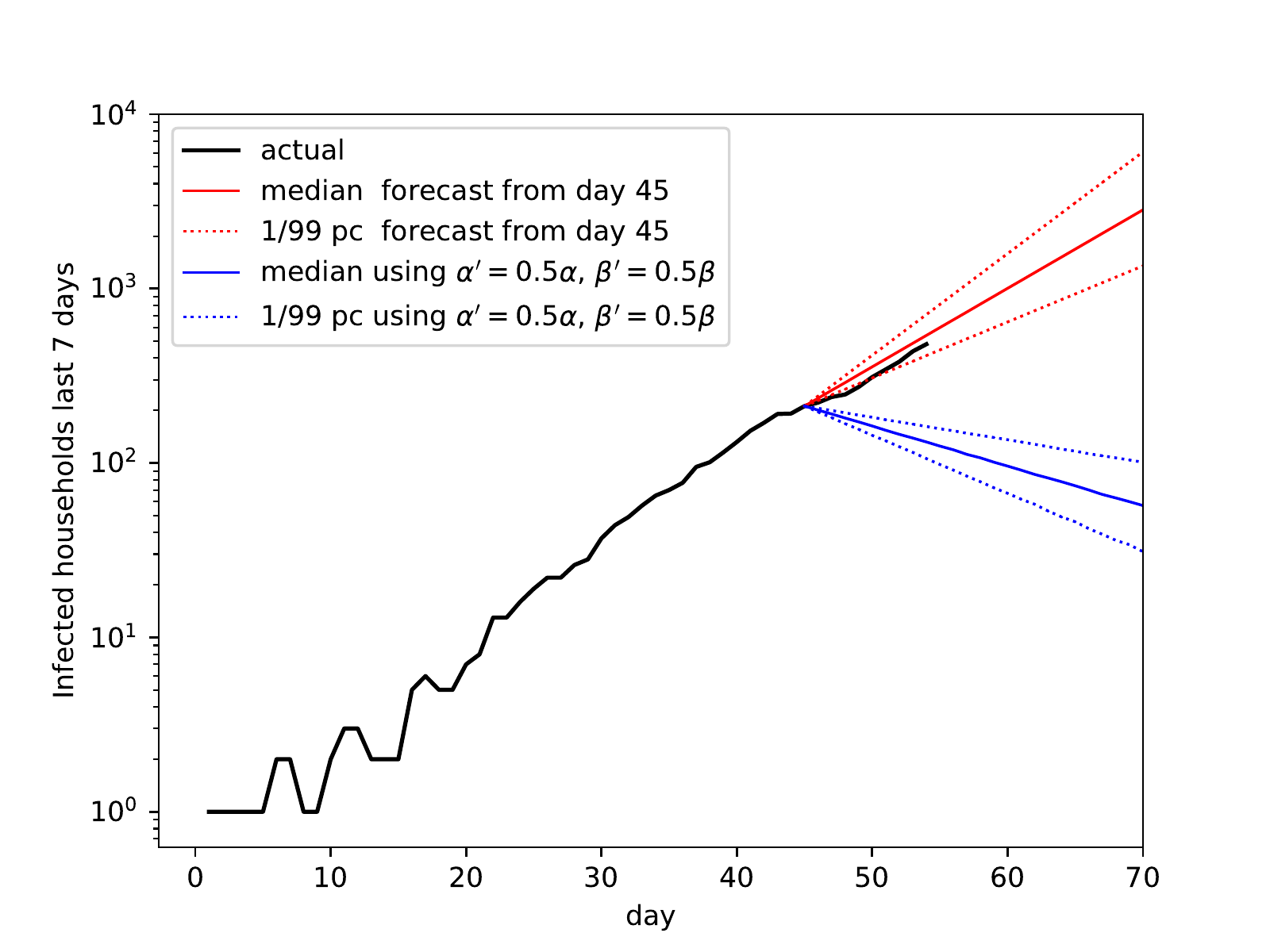}
  \caption{Sample forecasting results,
    with alternate forecast
    showing the effect of reducing
both the between-household transmission rate $\alpha$,
and the within-household transmission rate parameter $\beta$.}
\label{fig:forecast}
\end{figure}

In addition to forecasting with the inferred parameters for a given day,
forecasts may account for reduced infection rates,
re-calculating $r$ using the method in Section \ref{step3}.
An example is shown in Figure \ref{fig:forecast}.
The true (simulated) progress of the outbreak for the next 10 days is also included,
showing that the exponential approximation works reasonably well for a short-term forecast.

\section{Conclusion}\label{conclusion}

We have presented a method for inference of infection parameters
in a two-level infection model,
with different infection rates within households 
and between different households. 
By separating the parameters into two components,
we split a complicated inference problem into two relatively well understood problems.

The within-household infection parameters can be inferred using known
techniques such as a particle-marginal Metropolis-Hastings algorithm \cite{Schon2018, Walker2019}.
The inferred within-household parameters can then be used to estimate the effective
number of infector households on a given day, and hence estimate $\alpha$,
the between-household infection rate.
The technique is similar to other methods which infer the parameters dynamically during an outbreak \cite{Cori2013, Thompson2019},
but also uses knowledge of within-household structure and parameters,
and uses it to eliminate a systematic bias when the number of new infections per day is changing.

The within-household and between-household parameters can 
be combined, to give the reproduction number in a number of forms.
An exponential growth rate $r$ can also be estimated, and used for forecasting.
In this case, the required path integral can be efficiently estimated using simulation.
In all cases, the testing confirms that the method works well.

This work could be extended in a number of directions.
First, we emphasise that the method is not specific to any particular within-household model,
so can be adapted to a model for a different type of pathogen,
or a more or less detailed model of the internal structure of a household.
Also, the method scales well to the use of larger household groups;
unlike earlier methods \cite{Ross2010, Black2013}
which strained computer resources as household sizes reached even a moderate size.

For the between-household inference, we believe a household-based model
can be useful for estimating the spread of an infection through a population.
Our technique for overcoming the bias when the number of new infections per day is growing or decaying
(Section \ref{step2}) is not limited to a household-based model, so
may be applicable in other situations.
The technique presented here assumes a constant rate of change,
and it may also be useful to
refine this technique to account for changes in the rate of growth or decay.

There is also scope for extending the method for forecasting.
We present a very simple method, of calculating the exponential growth rate and
using that to project forward. But with all lower-level parameters inferred,
such as those relating to transmission of the infection and recovery of individuals,
it may be possible to make forecasts using those parameters directly,
without assuming a constant exponential growth rate.

In conclusion, we believe the stratification of data into households
is an effective method for inference of outbreak parameters,
and there is much scope for using these or similar methods in the future.

\section{Declaration of Competing Interest}

The authors declare that they have no known competing financial interests
or personal relationships that could have appeared to influence the work reported in this paper.

\section{Acknowledgements}

This material is based upon work supported by the
U.S. Army International Technology Center Pacific (ITC-PAC)
under Contract No. FA5209-17-C-0008.

This work was also funded by the Australian Defence Science and Technology Group
Bioterrorism Preparedness Strategic Research Initiative 17/491.

This work was supported with supercomputing resources provided by the Phoenix HPC service at the University of Adelaide.

Simulation and inference software was written in Julia \cite{Julia2017}, with post-processing in Python and Matplotlib \cite{Hunter2007}.

\section{References}\label{references}
\begingroup
\renewcommand{\section}[2]{}%

\bibliographystyle{plainnat}
\raggedright
\bibliography{biblio}

\endgroup

\section*{Appendices}
\appendix

\section{Conjugate prior proof}\label{conjproof}

%

Consider a set of event rates $\alpha \xi_1 \dots \alpha \xi_c$,
  where $\xi_1 \dots \xi_c$ are known but $\alpha$ is not known;
  and corresponding observed daily event counts $y = (y_1 \dots y_c)$.
  Given that the prior distribution of $\alpha$ 
is a gamma distribution with shape $a$ and rate $b$, meaning
\begin{equation}\label{prior}
  f(\alpha) = \frac{b^a}{\Gamma(a)} \alpha^{a-1}e^{-b\alpha} ,
\end{equation}
we wish to infer the posterior distribution $f(\alpha|y)$.


The number of daily events $Y_j$ is poisson distributed, with:
\begin{equation}\label{likeday}
  p(Y_j=y_j|\alpha) = 
  \frac{\left( \alpha \,\xi_j \right)^{y_j}
    e^{\left(-\alpha \,\xi_j\right)} } {y_j!} ;
\end{equation}
which means the likelihood $f(y|\alpha)$ is given by:
\begin{equation*}
  f(y|\alpha) = \prod_{j=1}^c p(Y_j=y_j|\alpha) 
  = \prod_{j=1}^c
  \frac{\left( \alpha \,\xi_j \right)^{y_j}
    e^{\left(-\alpha \,\xi_j\right)} } {y_j!}
\end{equation*}
\begin{equation}\label{likelong}
  =   \frac{\prod_{j=1}^c {\xi_j}^{y_j}}{\prod_{j=1}^c y_j!}
  \alpha ^{\left(\sum_{j=1}^c y_j \right)}
    e^{-\alpha \left( \sum_{j=1}^c \xi_j\right)} .
\end{equation}

Substituting \eqref{prior} and \eqref{likelong} into 
Bayes' rule, noting $\alpha > 0$, gives:
\begin{equation*}
  f(\alpha|y)
  = \frac{f(y|\alpha)f(\alpha)}{\int_0^{\infty}f(y|\alpha)f(\alpha)d\alpha}   
\end{equation*}
\begin{equation*}
  = \frac{
    \frac{ b^a \prod_{j=1}^c {\xi_j}^{y_j}}{\Gamma(a) \prod_{j=1}^c y_j!}
    \alpha^{(a-1+\sum_{j=1}^c y_j)} e^{-\alpha (b+\sum_{j=1}^c \xi_j)}
  }
       { \int_0^{\infty}
         \frac{ b^a \prod_{j=1}^c {\xi_j}^{y_j} }{\Gamma(a) \prod_{j=1}^c y_j!}
          \alpha^{(a-1+\sum_{j=1}^c y_j)} e^{-\alpha (b+\sum_{j=1}^c \xi_j)} d\alpha
         } .
\end{equation*}

To simplify the logic, we define
\begin{equation}\label{SR}
S=a+\sum_{j=1}^c y_j \quad \text{ and } \quad R=b+\sum_{j=1}^c \xi_j ,
\end{equation}
giving:
\begin{equation*}
  f(\alpha|y) =
  \frac{ \alpha^{S-1} e^{-R \alpha} }
  { \int_0^{\infty} \alpha^{S-1} e^{-R \alpha} d\alpha } 
\end{equation*}
\begin{equation}\label{ND2}
  =  \frac{\frac{R^S}{\Gamma(S)} \alpha^{S-1} e^{-R \alpha} }
  { \int_0^{\infty} \frac{R^S}{\Gamma(S)} \alpha^{S-1} e^{-R \alpha} d\alpha } .
\end{equation}

The quantity inside the integral in 
\eqref{ND2} is a gamma distribution,
meaning it integrates to $1$, giving:
\begin{equation*}
   f(\alpha|y) =
   \frac{R^S}{\Gamma(S)} \alpha^{S-1} e^{-R \alpha} ,
\end{equation*}
which is itself a gamma distribution, with shape $S$ and rate $R$; that is:
\begin{equation}\label{post}
  f(\alpha|y) \sim
  \text{Gamma}\left(\text{shape}=a+\sum_{j=1}^c y_j, \quad \text{rate}= b+\sum_{j=1}^c \xi_j \right) .
\end{equation}
which is equivalent to \eqref{post1}.
\qed
%

\section{Inference during growth and decay}\label{Keffect}

  This appendix concerns the correction to account for the growth or decay of an outbreak,
which is discussed in Section \ref{fixed},
and appears as the 
$K$ term in \eqref{xi}, \eqref{xim} and \eqref{xim_indiv}.
We illustrate how, without this term,
we see a bias due to the growth or decay of the outbreak,
as has previously been reported \cite{ScaliaTomba2010, Britton2019, Champredon2015};
and that this $K$ term provides an appropriate correction.

\begin{figure}[H]
\begin{subfigure}{.8\textwidth}
  \centering
  \includegraphics[width=1\linewidth]{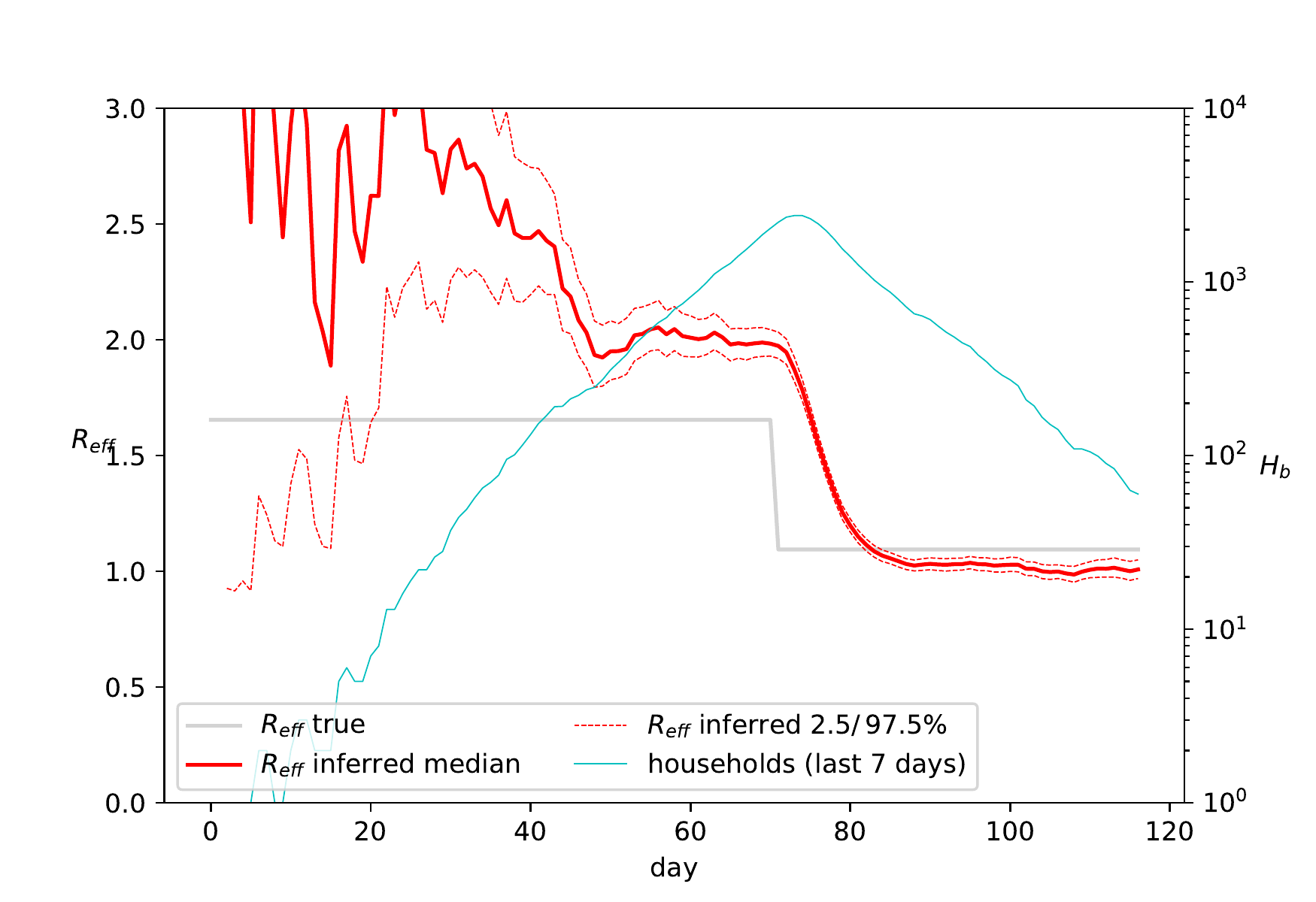}
  \caption{without correction}
  \label{app_r0_K1}
\end{subfigure}
\begin{subfigure}{.8\textwidth}
  \centering
  \includegraphics[width=1\linewidth]{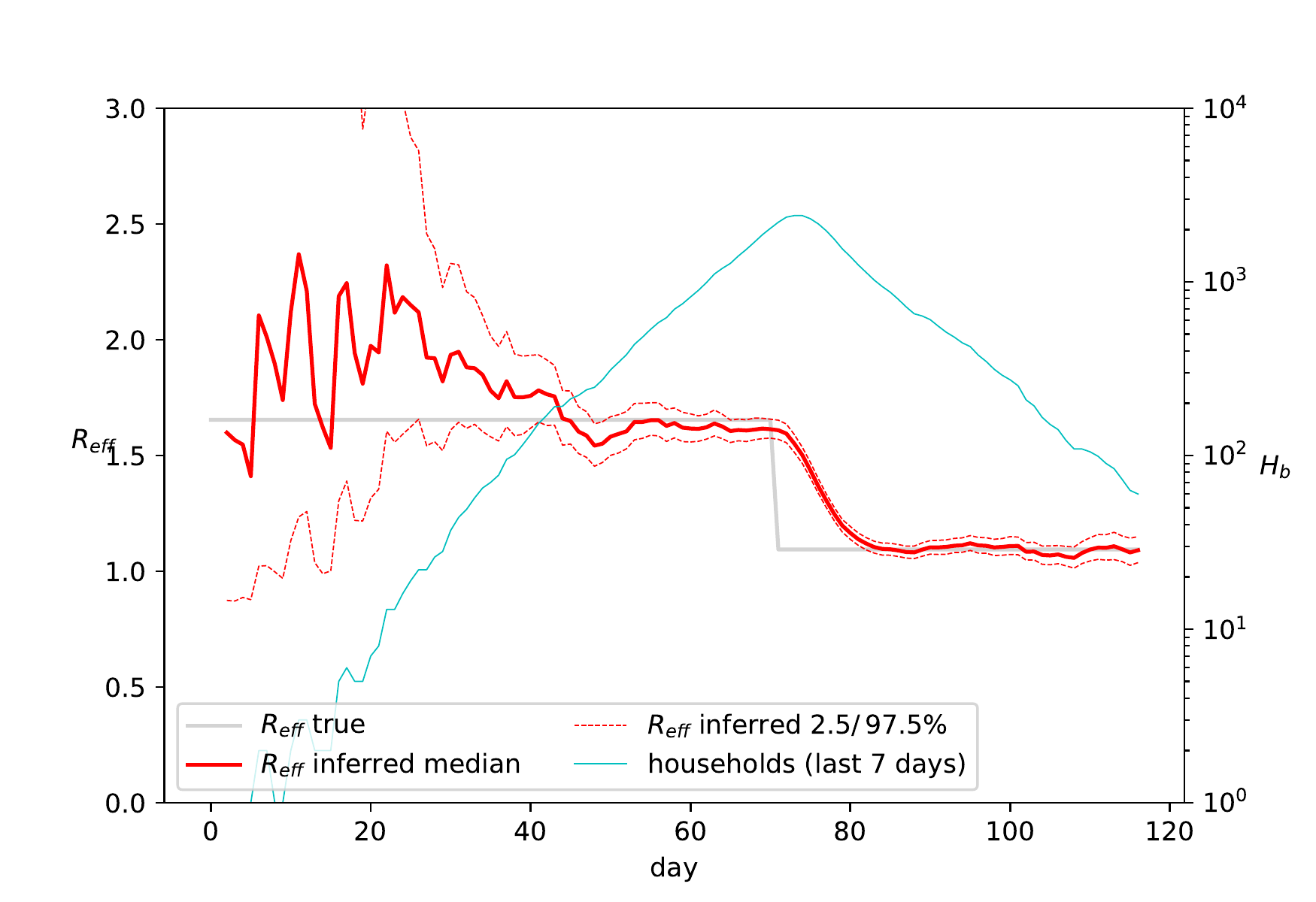}
  \caption{with correction} 
  \label{app_r0}
\end{subfigure}
\caption{Inference of $R_\text{eff}$
  during a simulated outbreak
  between-household inference updated daily, using the most recent $7$ days of data;
  and the within-household parameters known exactly.
  Sub-figure (a) omits the correction due to growth or decay,
  by setting $K_{(k,m)}=1$ in place of \eqref{Km}.
  Sub-figure (b) uses the method discussed in the text.
  The dotted red lines show the $2.5$ and $97.5$ percentiles of the posterior;
  the true value of $R_\text{eff}$ is shown in grey;
  and the number of households infected in the most recent 7 days ($H_b$) is shown in light blue.}
  \label{Kdemo}
\end{figure}

\begin{figure}[H]
\begin{subfigure}{.8\textwidth}
  \centering
  \includegraphics[width=1\linewidth]{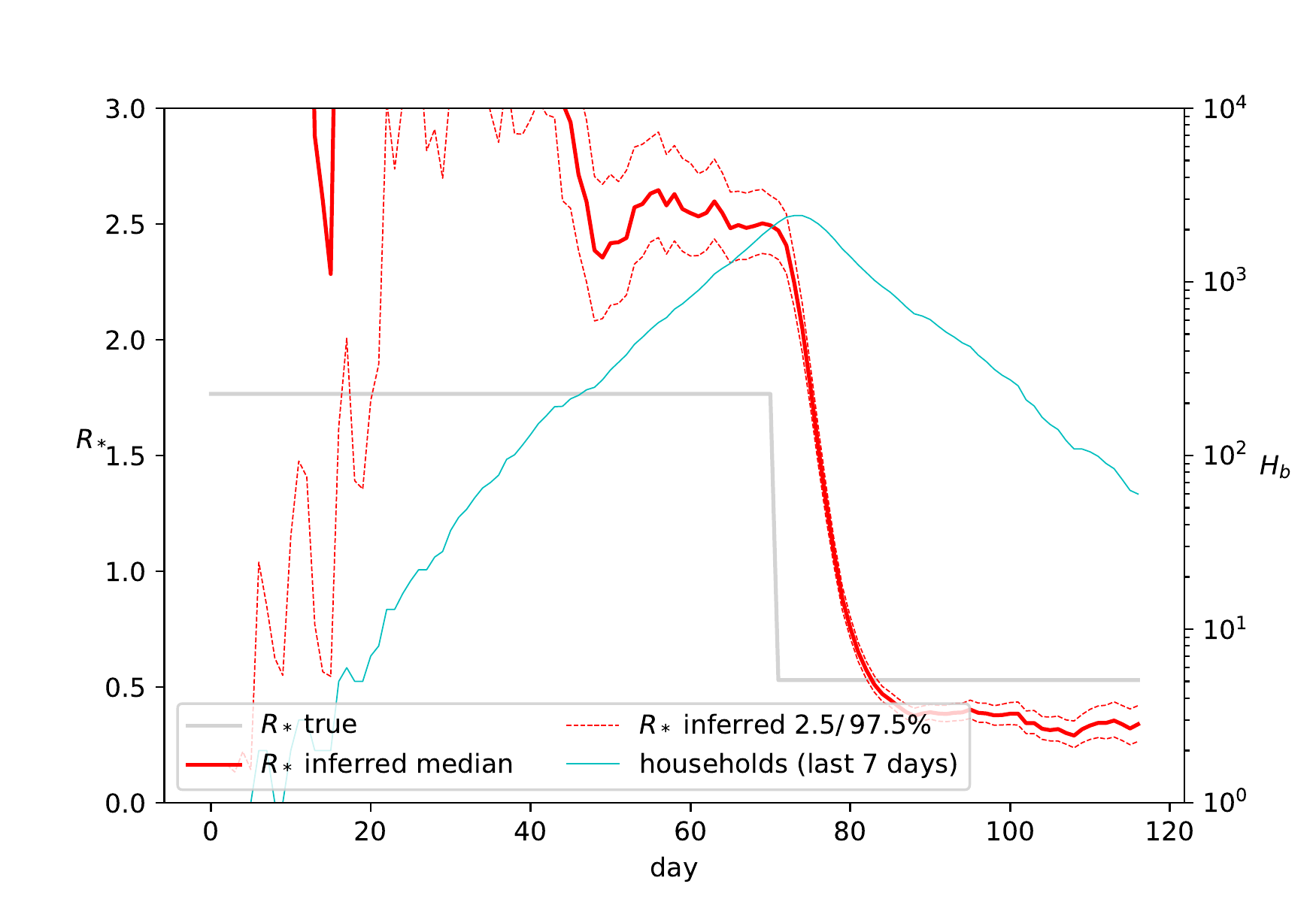}
  \caption{without correction}
  \label{app_rstar_K1}
\end{subfigure}
\begin{subfigure}{.8\textwidth}
  \centering
  \includegraphics[width=1\linewidth]{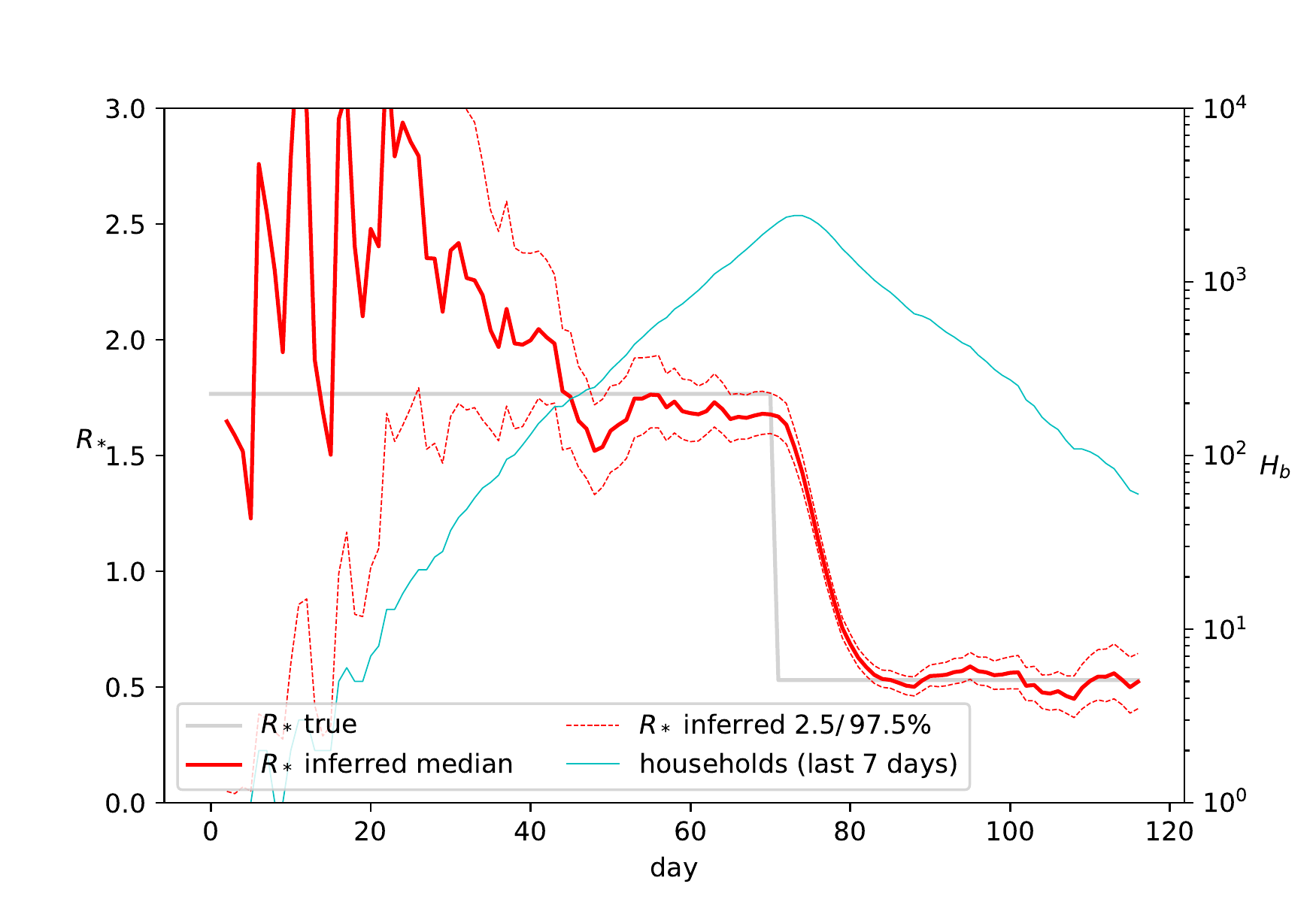}
  \caption{with correction} 
  \label{app_rstar}
\end{subfigure}
\caption{Inference of $R_*$
  during a simulated outbreak
  between-household inference updated daily, using the most recent $7$ days of data;
  and the within-household parameters known exactly.
  Sub-figure (a) omits the correction due to growth or decay,
  by setting $K_{(k,m)}=1$ in place of \eqref{Km}.
  Sub-figure (b) uses the method discussed in the text.
  The dotted red lines show the $2.5$ and $97.5$ percentiles of the posterior;
  the true value of $R_*$ is shown in grey;
  and the number of households infected in the most recent 7 days ($H_b$) is shown in light blue.}
  \label{Kdemostar}
\end{figure}

Figures \ref{Kdemo} and \ref{Kdemostar} show inference results,
of $R_\text{eff}$ and $R_*$ respectively,
from the same simulation. 
This simulation uses the
distribution of household sizes given in Table \ref{tab:pop};
within-household parameters
$t_E=2$, $t_P=1.8$, $t_I=1.5$, $p_s=0.8$,
$R_{0i}=\beta(t_P+t_I)=1.4$;
and $\alpha$ initially $0.242$, then reducing to $0.0727$ on day $70$.
For inference, the within-household parameters are assumed to be known exactly,
while $\alpha$ has a gamma distribution with shape $1$ and scale $0.3$.
That is, all parameters and settings are identical to those used for Figure \ref{time_fixed},
except that $\alpha$ reduces more sharply,
to exaggerate the rate of decay of $R_\text{eff}$ and $R_*$.

Without the correction of the $K$ term, the inference method
overestimates $\alpha$ when the outbreak is growing,
and underestimates $\alpha$ when the outbreak is decaying,
causing bias in $R_\text{eff}$ (Figure \ref{Kdemo}(a)) and $R_*$ (Figure \ref{Kdemostar}(a)).

When the $K$ term is included,
the inference accounts for this growth and decay,
as is illustrated in Figures \ref{Kdemo}(b) and Figure \ref{Kdemostar}(b) respectively.

\end{document}